\documentstyle[11pt,aaspp4]{article}

\newcommand{\lya}{Ly$_\alpha$\ }
\newcommand{\ha}{H$_\alpha$\ }
\newcommand{\hb}{H$_\beta$\ }
\newcommand{\oii}{[O{\small II}]$\lambda$3727\ }
\newcommand{\Oii}{[O{\small II}]}

\newcommand{\Oiii}{[O{\small III}]}
\newcommand{\CGS}{\,$\times 10^{-16}$\,erg cm$^{-2}$ sec$^{-1}$}
\newcommand{\cgs}{\,$\times 10^{-16}$ cgs}
\newcommand{\cgst}{$10^{-16}$ cgs}
\newcommand{\sfr}{\,M$_{\odot}$/yr}

\newcommand{\tm}{$\times$}

\slugcomment{}

\lefthead{Mannucci et al.}
\righthead{Infrared emission-line galaxies}

\begin{document}


\title{Infrared emission-line galaxies associated with damped \lya 
and strong metal absorber redshifts 
\footnote{Based on observations collected with
the DSAZ telescopes in Calar Alto and ESO telescopes in La Silla}}

\author{F. Mannucci}
\affil{C.A.I.S.M.I.--C.N.R., Florence, Italy\\
Electronic mail: filippo@arcetri.astro.it}

\author{D. Thompson \& S. V. W. Beckwith}
\affil{Max-Planck-Institut f\"ur Astronomie, Heidelberg, Germany\\
Electronic mail: djt@mpia-hd.mpg.de, svwb@mpia-hd.mpg.de}

\and

\author{G. M. Williger} 
\affil{Goddard Space Flight Center, Greenbelt, MD\\
Electronic mail: williger@tejut.gsfc.nasa.gov}

\begin{abstract}
Eighteen candidates for emission line galaxies were discovered 
in a narrow-band infrared survey that targeted the redshifts of 
damped \lya or metal lines in the spectra of quasars.  
The presence of emission lines is inferred from the photometric 
magnitudes in narrow and broad band interference filters, corresponding
to \ha at redshifts of 0.89 (6 objects) and 2.4 (10 objects),
and \oii at a redshift of 2.3 (2 objects).  
Most of the candidates are small resolved objects, 
compatible with galaxies at the redshifts of the absorbers.
Because a similar survey targeted at the redshifts of
quasars themselves uncovered only one emission-line galaxy 
in a larger volume, the results imply substantial clustering of young
galaxies or formation within filaments or sheets whose locations are
indicated by the redshifts of absorption along the
lines of sight to more distant quasars.
\end{abstract}

\vspace*{2mm}


\newpage

\section{Introduction}

Exploration of the high redshift universe and discovery of the most distant
objects is still in its infancy.  
Only recently have the tools been available
to detect normal galaxies at redshifts larger than one, when the first
galaxies were created (\cite{pescarelle96,hu96,cowie98,steidel96}).  
It seems likely that young galaxies will have
a variety of different signatures so that it will be necessary to use
several diverse techniques to uncover all of them: searches at optical,
infrared, x-ray, and radio wavelengths, for example.
In particular, basing the statistical studies of the high redshift
galaxies only on objects detected in the rest-frame UV could miss many
young galaxies (\cite{franceschini98,guiderdoni97}), 
and sampling longer wavelength ranges is necessary.

We carried out a survey for infrared emission-line galaxies by
imaging through narrow (${\Delta\lambda \over \lambda} \sim 0.01$) and
broad band filters between 1 and 2.5\,$\mu$m, identifying objects
that appeared brighter in the narrow filters
(\cite[hereafter TMB96]{tmb96}). Our first survey was designed to uncover
emission lines at the redshifts of quasars within each survey field, in
case there is substantial clustering marked by quasars.
In an area of 276\,\sq\arcmin , only one
emission-line galaxy was discovered (\cite{beckwith98}).  The
surface density of such objects implied by these results is similar to
that inferred from other surveys (\cite{cowie94,graham96,malkan96,bechtold97}) 
and suggests that the infrared emission-line galaxies constitute at most a 
modest population of young galaxies at high redshift.

Using the same instruments, we undertook a second infrared survey for 
emission-line galaxies targeted at the redshifts of damped
\lya or metal absorption lines 
in the spectra of quasars.  Damped \lya absorbers are thought to
contain as much baryonic matter as seen in all spiral galaxies today
(\cite{wolfe86}) and may, therefore, mark sites of ongoing star
formation.  
Several other groups 
(\cite{lowenthal91,macchetto93,wolfe92,moller93,djorgovski96,francis97}) 
carried out similar surveys at optical wavelengths 
looking for \lya emission-line galaxies in these regions.
They discovered only a few such emission-line (non-AGN) galaxies, but
\cite{wolfe93} showed that the implied volume density was significantly 
higher than in the general field.
Metal absorption systems also indicate that star formation
has taken place, and are identifiable from the ground at lower
redshifts than \lya alone.

We selected damped \lya systems or metal absorbers whose redshifts put 
the main optical lines H$_\alpha$, H$_\beta$, \Oiii, and \Oii\ into 
standard narrow-band filters in the J, H and K
bands.  The resulting redshift ranges are: $0.5<z<1.9$, $2.1<z<2.5$
and $3.1<z<3.8$.  Special emphasis was given to the \ha\ line 
expected to be the brightest in young star systems and the least affected
by dust.  
This letter describes the results of the new survey.

\section{Observations}

As described in TMB96, pairs of narrow and broad-band images were taken
of the selected fields.
Most of the data, 163\,\sq\arcmin\ in 13 fields, were obtained at the 
Calar Alto 3.5m telescope, using the NICMOS3 256$^2$ MAGIC 
cameras (\cite{herbst93}) with a resolution of 0\farcs 81 per pixel. 
One field, 38.6\,\sq\arcmin, was obtained at the same telescope with the 
Omega Prime camera using a 1024$^2$, HgCdTe Hawaii array with 0\farcs 40 
per pixel. Five more fields for a total of 26.2\,\sq\arcmin\ were observed 
with the IRAC2b camera at the 2.2m ESO/MPI telescope at La Silla.  
The area-weighted average limiting flux is 2.4\CGS, or 1.6\CGS, if only
the Calar Alto data are considered.
The comoving volume sampled by this survey at the redshift of the
absorbers is about 20,600\,Mpc$^3$, assuming only the target line at
the appropriate redshift could be detected.  
Considering all the four principal optical lines \Oii, \hb, \Oiii\ and \ha ,
the total sampled volume is 90,000\,Mpc$^3$ 
($H_0 = 50$, $q_0 = 0.5$, assumed throughout this paper).
For comparison, the total sampled volume by the first survey 
targeting quasars redshifts is 153,000\,Mpc$^3$.

\section{Results}

\begin{figure}
\plotone{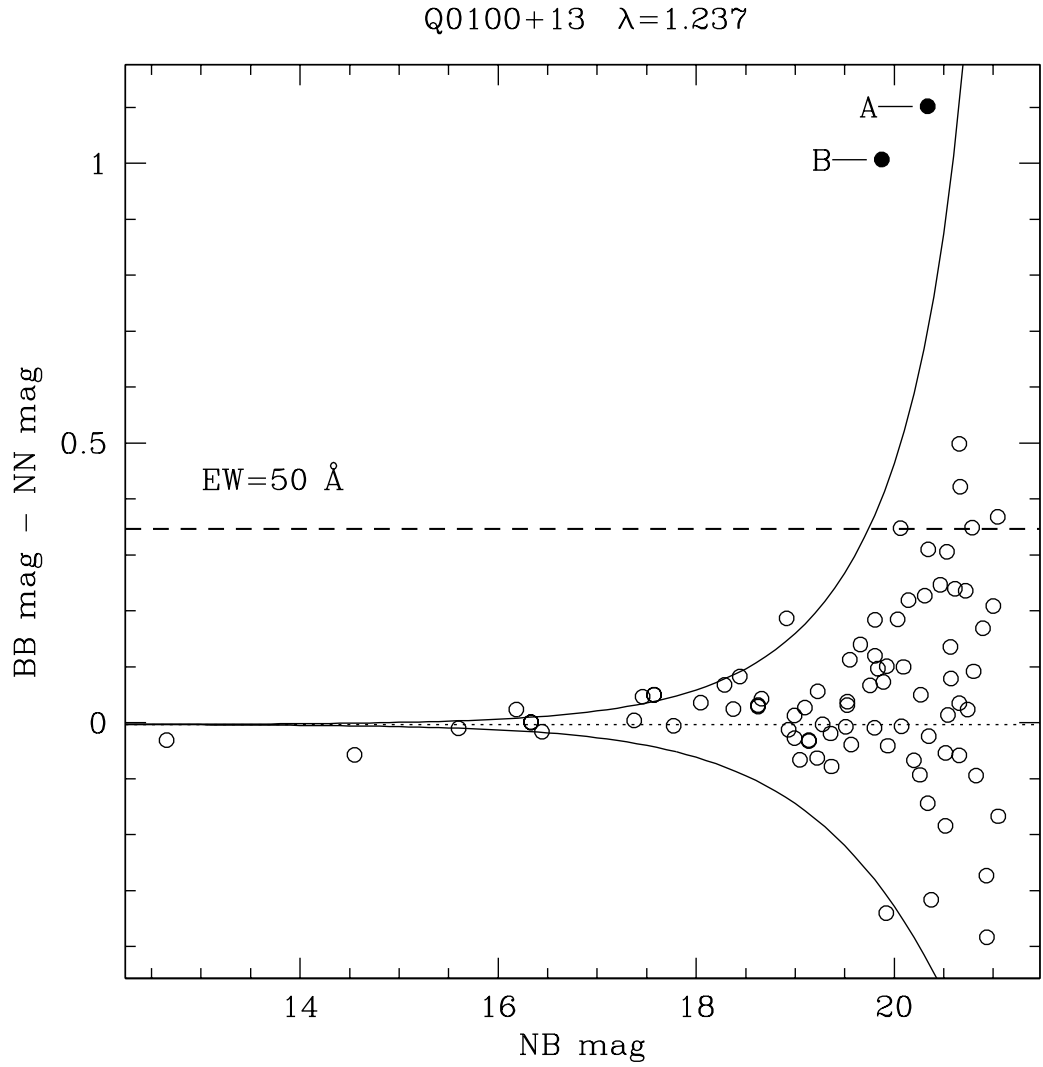}
\figcaption[Mannucci.fig1.ps]{Color -- magnitude diagram for the 
Q0100+130 (PHL 957) field.  The solid lines
indicate the 3$\sigma$ uncertainties, the dashed line shows the
position of objects with an emission line with EW=50\AA. The 
two candidate emission-line galaxies are marked as A and B.
\label{fig1}}
\end{figure}

Objects were selected if well detected in the narrow-band images, with
narrow-band magnitudes exceeding the broad-band magnitudes by more than 2.5
standard deviations of the combined uncertainties in the two filters,
and with derived
line equivalent width larger than 50\,\AA .  These criteria are
easily checked by plotting objects in a color-magnitude diagram,
where the color is the broad minus narrow band magnitude.
Figure~\ref{fig1} is a color-magnitude diagram of the Q0100+130 (PHL~957) field
showing the detection of two candidates discovered in the J band.
The candidates, denoted A \& B, clearly stand out above the noise 
at high equivalent width, although A is near the narrow-band limit.
This field also contains a \lya emission-line galaxy discovered 
by \cite{lowenthal91} and subsequentely imaged in the H$\alpha$+[NII] lines
at 2.177\,$\mu$m by \cite{bunker95}.
This object is not seen in our image, and 
its expected \oii flux (about 0.7 \cgs) derived from the H$\alpha$ flux
by \cite{bunker95} is below our detection threshold (about 1.4 \cgs).
Neither A nor B is seen in the \cite{lowenthal91} image, while
our candidate A is barely visible
on the \cite{bunker95} narrow band image, implying faint H$\alpha$
emission.  Both A and B are visible in their broad band K image, as
well, thus supporting the reality of the detections in our survey.

From the color-magnitude diagrams of five of the 19 fields in this survey,
we discovered 18 candidates for emission-line galaxies. 
The emission lines, if spectroscopically confirmed, would correspond
to \ha at redshifts of 0.89 (6 objects) and 2.4 (10 objects),
or \oii at a redshift of 2.3 (2 objects).  
Most of the objects are a few seconds of arc in extent suggesting that 
they are galaxies at redshifts greater than a few tenths.

Table~1 lists the coordinates of the candidates, their offsets from the 
quasar, and their morphology as deduced from our images.  The angular
distances from the quasars are between 9\arcsec\ and 120\arcsec ,
corresponding to projected distances between 70\,kpc and 1\,Mpc at the
distances to the absorption-line systems.  Very few, if any, of these
objects could, therefore, be identified with the absorption-line
objects, whose typical extents are probably those of
galaxies or protogalactic clumps, $\sim 5 - 60$\,kpc across 
(\cite{fukugita96,haehnelt96}).
However, due to the coarse sampling in our images, no attempt was made yet 
to subtract a PSF from the QSO
image, so objects within a few arcsec from the QSO would not
have been easily seen.


Table~2 gives the magnitude, a rank of the significance of the
detection, the statistical signal-to-noise ratio of the emission line,
and the line flux for each candidate.  
The final four columns give the line identification, redshift, rest
equivalent width, and derived star formation rate (\cite{mannucci95}; TMB96)
assuming the line is at the redshift of the absorption-line.  
The derived SFRs attribute all of the line emission
to H{\small II} regions, ignoring any contribution from an active
galactic nucleus (AGN), an assumption that may be incorrect in at least some
cases (\cite{beckwith98}).

\begin{figure}
\plotone{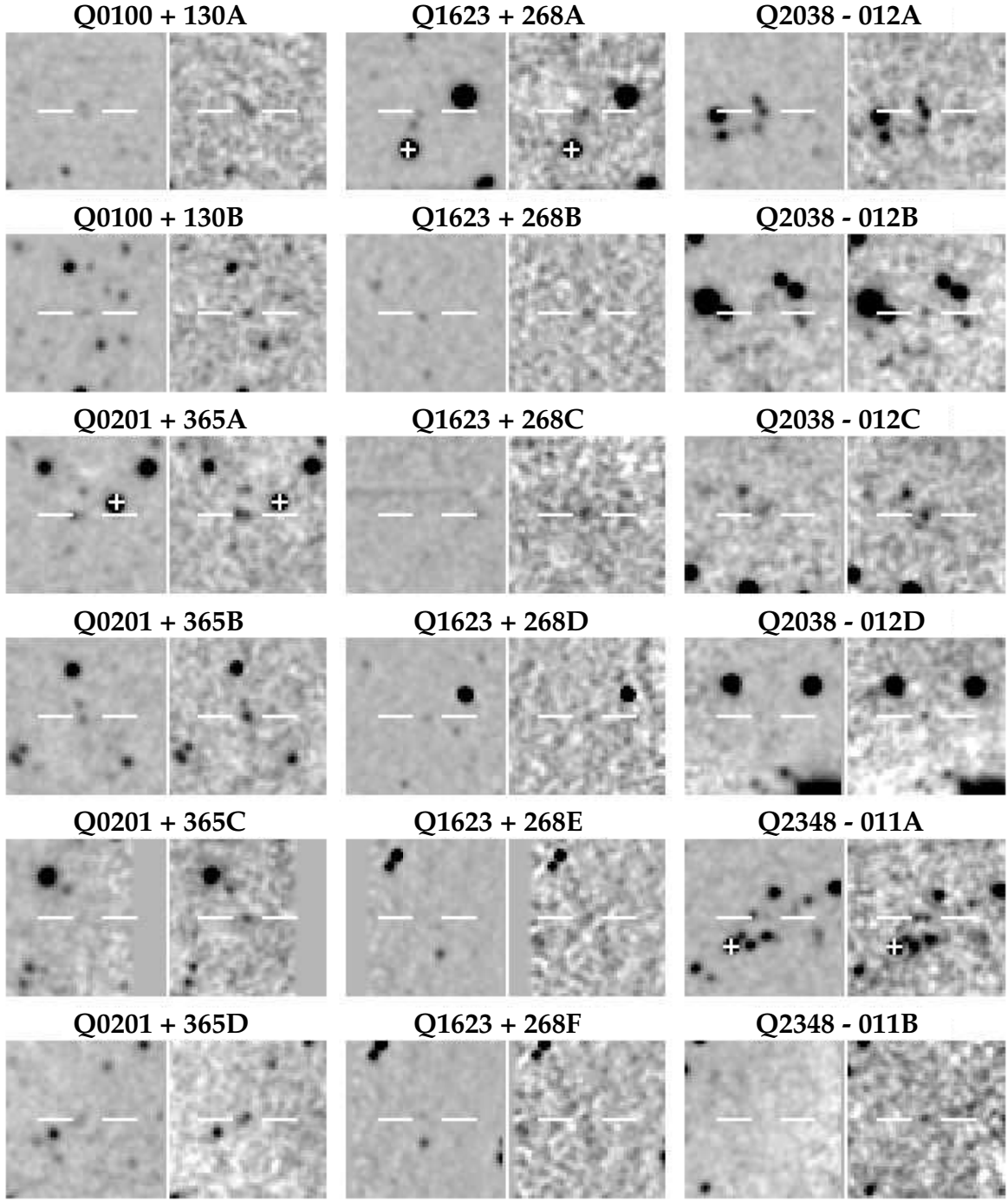}
\figcaption{Images of the candidates. In each panel, the BB image is on the
left and the corresponding NB image on the right, with the
candidate at the center. A white cross marks the QSO when present in the
field. The images have been convolved with a gaussian with FWHM about
equal to the seeing to make the faint objects more visible.
The dimensions of each subimage are 40 arcsec, North is up and East left.
\label{fig2}}
\end{figure}
\begin{figure}
\plotone{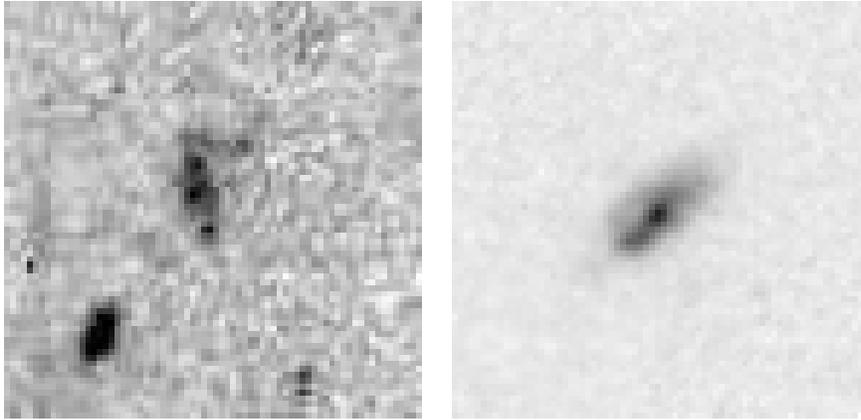}
\figcaption{HST WFPC2 images of two of the candidates in the 
Q1623+268 field: A (left) and D (right).  
The dimensions of each image are 72 pixels or 7.2 arcsec on each side
The images have been rotated from the original WFPC2 orientation so
that north is up and east to the left.
\label{fig3}}
\end{figure}

Figure~\ref{fig2} shows images of each candidate in the narrow and broad 
filters.
The typical angular resolution is about 1\farcs 2, corresponding 
to about $\sim 10$\,kpc for redshifts between 1 and 3.  Most of the
objects appear resolved even at  this resolution. 
The field around Q1623+268 was observed by Steidel in 1996 May with
the Hubble Space Telescope using the WFPC2 camera with the 
F702W filter and total integration time of 87 min.
Two of our emission-line candidates, Q1623+268A and Q1623+268D, 
appear in these images, parts of which are shown in Figure~\ref{fig3}.  
Both objects appear to be late-type spiral or irregular galaxies.
The object sizes are 2\farcs 4 (Q1623+268A) and 3\farcs 6
(Q1623+268D), corresponding to 20 and 30 Kpc for $z\simeq0.9$. If these
objects are at the assumed redshifts, they are large, well formed
galaxies. The detected star formation activity with SFRs between 5 and
10 \sfr\ would be the normal activity of late-type spiral or irregular
galaxies.

\section{Discussion}

The most striking result of this survey is the large number of candidates
discovered in a small area implying that by choosing the right redshift
intervals, it is possible to detect emission-line objects rather easily.  
The total sky coverage in this survey was 228\,\sq\arcmin\ compared
with 276\,\sq\arcmin\ in TMB96, yet 18 candidates were discovered in
the second survey compared to a single emission-line galaxy in the
first.  Taken at face value, the new results suggest that damped
\lya and metal absorbers pinpoint the regions where galaxy formation
is taking place.

Confirmation of the exact nature of these objects will require
spectroscopic follow up and perhaps additional imaging with HST.
Since not all the objects are resolved at the
coarse pixel scale used, some could be stars or other
nearby objects.  A few objects are near the 3$\sigma$ detection limit.
Statistical comparison must, therefore, be regarded with some caution
until further data are available.

Nevertheless, we believe that the main
result of this survey is robust: there are more emission-line
galaxies associated with damped \lya and metal 
absorber redshifts than with the 
redshifts of quasars or at arbitrary redshifts.  First, the observational
method employed was identical to that of TMB96, but the resulting
number of candidates is almost 20 times larger. 
Second, the one candidate, cK39, discovered in the first survey by
TMB96 was also coarsely sampled
but is, indeed, an emission line galaxy (\cite{beckwith98})
spectroscopically confirmed and easily resolved with the Keck telescope.  
Third, the HST images of two of our {\em least} statistically significant
candidates show them to be galaxies of exactly the size expected for 
objects at redshifts greater than a few tenths and with the kind of 
morphology expected of objects in late stages of assembly.  

These results may be compared with similar studies of \lya emitting
galaxies in the neighborhood of damped \lya absorption line systems.
By the number of detected objects 
(\cite{lowenthal91,wolfe92,macchetto93,moller93,djorgovski96})
and observed fields (\cite{smith89,deharveng90,lowenthal95})
an average comoving density of these systems of about
7$\times10^{-4}$ Mpc$^{-3}$ is derived down to a limiting SFR 
of about 5\,\sfr, assuming case B recombination and no dust.
The infrared technique gives similar results, i.e., detections in about 
1/4 of the fields and a density of $9\times 10^{-4}$\,Mpc$^{-3}$. 
The range of SFR implied by our observations
6 - 200\,\sfr , is somewhat higher than that usually obtained by the
optical searches, 5 - 20\,\sfr , but not dramatically so.  
If there is modest local extinction to the star formation regions in
high redshift galaxies, optical derived SFRs 
must be increased by a factors
of about 3 to get the true SFR (\cite{pettini97}), making their range
almost coincident with ours.
Our results are
perhaps more consistent with the limiting SFR of 30 - 80\,\sfr\
along sightlines toward eight $z>2$ 
damped \lya absorbers as derived from redshifted H$\alpha$
(\cite{bunker98}). 

On the other hand, the lines could be produced by active galactic nuclei.
There are several good reasons to believe this could be the case for the one
emission-line object discovered in the survey of TMB96 (\cite{beckwith98}),
and that it could be a widespread phenomenon (\cite{francis97}).
If so, the density of such objects in these regions is considerably higher
than the cosmic average.  The density of known quasars at these redshifts 
(e.g., \cite{andreani92}) implies about $4-8\times10^{-5}$  
quasars per field, meaning that we might be seeing far more AGNs than expected.
Although the damped \lya redshifts mark regions
with many galaxies, the nature of each single galaxy will remain a 
mystery until proper follow up observations will be able to
distingish between emission due to star formation and AGNs.

Cold dark matter simulations of early galaxy formation produce
filaments of galaxies (\cite{white94}) with groups of high redshift
metal and damped \lya absorbers spanning up to several hundred kpc
(\cite{rauch97}).  If the damped \lya absorbers
actually trace the positions of these filaments, then the high
detection rate of emission line galaxies in this survey may support
the overall structure predicted by CDM.  This conclusion can be 
made quantitative by measuring the average overdensity of objects 
near damped \lya and metal systems with respect to TMB96.\\

\vspace*{3mm}

\acknowledgments

We are grateful to the staffs at the Calar Alto 
Observatory and ESO at 
La Silla for excellent assistance with the observations.  We profited 
from discussions with Esther Hu, Garth Illingworth, Matt Malkan, 
Ed Salpeter and Chuck Steidel.
This research was supported by the Max-Planck-Society and also uses
Archival data from the Hubble Space Telescope.  
FM acknowledges partial support by ASI grant ARS-96-66.\\


\clearpage

\makeatletter

\begin{deluxetable}{lccccccl}
\tablewidth{0pt}
\tablenum{1}
\tablecaption{The Emission-line Candidates \label{table1}}
\tablehead{
	\colhead{Object} 		&
	\colhead{RA(2000)}		&  
	\colhead{DEC(2000)}		& 
	\colhead{$\Delta$\tablenotemark{a}}	& 
	\colhead{D\tablenotemark{b}}			&  
	\colhead{P.A.\tablenotemark{b}}			&  
	\colhead{Size\tablenotemark{c}}		&  
	\colhead{Shape} 		\\
	\colhead{} 				&
	\colhead{}				&  
	\colhead{}				& 
	\colhead{($\prime\prime$)}		& 
	\colhead{($\prime\prime$)}		&  
	\colhead{(deg.)}		&  
	\colhead{($\prime\prime$)}		&  
	\colhead{}				\\
}
\dummytable\label{tab1}
\startdata
Q0100+130A & 01:03:10.39 &+13 17 03.7 & 1.1&  48 &344.3 & 5.4\tm 1.5& possibly double \\ 
Q0100+130B & 01:03:13.30 &+13 16 58.7 & 0.5&  51 & 35.3 & 1.9\tm 1.2& unresolved \\ 
\hline
Q0201+365A & 02:04:56.28 &+36 49 14.9 & 0.5&   9 &112.6 & \tablenotemark{d} & unresolved \\ 
Q0201+365B & 02:04:56.86 &+36 50 02.2 & 0.5&  46 & 18.9 & 1.8\tm 1.8& resolved \\ 
Q0201+365C & 02:04:50.20 &+36 47 47.4 & 0.5& 112 &215.6 & 2.4\tm 1.4& resolved \\ 
Q0201+365D & 02:05:01.75 &+36 47 47.0 & 0.5& 117 &141.0 & 1.4\tm 1.2& core/asymm. halo \\ 
\hline
Q1623+268A & 16:25:48.56 &+26 47 07.9 & 0.5&   9 &340.9 & 1.6\tm 1.0& resolved \\ 
Q1623+268B & 16:25:51.30 &+26 47 21.1 & 0.5&  40 & 56.7 & 3.8\tm 2.2& irregular, diffuse \\ 
Q1623+268C & 16:25:51.62 &+26 48 17.2 & 0.5&  87 & 25.9 & 2.5\tm 2.5& irregular \\ 
Q1623+268D & 16:25:43.39 &+26 45 40.9 & 0.5& 106 &222.8 & 2.9\tm 1.8& faint, elongated \\ 
Q1623+268E\tablenotemark{e}& 16:25:56.69 &+26 46 08.1 & 0.5& 117 &115.6 & 1.1\tm 1.1& faint, unresolved \\ 
Q1623+268F\tablenotemark{e}& 16:25:56.42 &+26 46 06.1 & 0.5& 152 &117.3 & 1.8\tm 1.0& core/asymm. halo \\ %
\hline
Q2038$-$012A & 20:40:52.99 &$-$01 05 29.2 & 0.7&  25 & 69.6 & 2.5\tm 1.5& elongated \\ 
Q2038$-$012B & 20:40:47.62 &$-$01 07 15.8 & 0.7& 113 &210.0 & 1.5\tm 1.5& diffuse \\ 
Q2038$-$012C & 20:40:53.61 &$-$01 07 32.6 & 0.7& 119 &163.7 & 2.2\tm 1.2& irregular \\ 
Q2038$-$012D & 20:40:57.93 &$-$01 06 50.8 & 0.7& 122 &126.5 & 1.3\tm 1.2& unresolved \\ 
\hline
Q2348$-$011A & 23:50:57.29 &$-$00 52 02.5 & 0.7&  11 &225.6 & \tablenotemark{d}      & unresolved \\ 
Q2348$-$011B & 23:50:54.64 &$-$00 52 58.4 & 0.7&  68 &315.6 & 1.6\tm 1.2& faint, elongated \\ 
\tablenotetext{a} {Radial uncertainty in the position}
\tablenotetext{b} {Projected distance and position angle (north to
east) from the QSO}
\tablenotetext{c} {Object sizes measured by the FWHM along respectively 
	the major and minor axis of elliptical gaussians fitted to the objects}
\tablenotetext{d} {not fitted because of the presence of a nearby object}
\tablenotetext{e} {the distance between objects is only 4\arcsec\ (33 Kpc); 
	they could be one galaxy}
\enddata
\end{deluxetable}

\clearpage

\begin{deluxetable}{lccrccccccc}
\tablewidth{0pt}
\tablenum{2}
\tablecaption{Candidate Magnitudes and Derived Properties \label{table2}}
\tablehead{
	\colhead{Object}        & 	
	\colhead{$\lambda$(NB)}& 	
	\colhead{NB}  &    	
	\colhead{BB}~~&	
	\colhead{R\tablenotemark{a}}&	
	\colhead{S/N}&
	\colhead{Line flux}&Line& 
	\colhead{z} &
	\colhead{EW$_{r}$} &
	\colhead{SFR}\\
	\colhead{}&  
	\colhead{$\mu$m}\ \ \ &
	\colhead{(mag)}&
	\colhead{(mag)}&
	\colhead{}&
	\colhead{}&
	\colhead{(\cgst)}  &
	\colhead{}&
	\colhead{}&
	\colhead{(\AA)} &
	\colhead{(\sfr)}
}
\dummytable\label{tab2}
\startdata
Q0100+130A    &1.237&20.33&   21.44& 2&3.8&1.9$\pm$0.5&\Oii&2.31&75$\pm$26&198$\pm$50\\
Q0100+130B    &1.237&19.88&   20.88& 2&5.6&2.7$\pm$0.5&\Oii&2.31&63$\pm$14&283$\pm$50\\
\hline
Q0201+365A\tablenotemark{b}&2.248&18.45&   19.37& 2&4.2&1.8$\pm$0.4&\ha &2.42& 65$\pm$20&68$\pm$16\\
Q0201+365B    &2.248&18.55&   19.62& 2&4.3&1.8$\pm$0.4&\ha &2.42& 88$\pm$27&71$\pm$16\\
Q0201+365C    &2.248&18.92&$>$20.78& 2&4.2&1.9$\pm$0.4&\ha &2.42& $>$182 &58--92\\
Q0201+365D    &2.248&18.71&   20.12& 1&4.5&2.0$\pm$0.4&\ha &2.42&158$\pm$62&76$\pm$16\\
\hline
Q1623+268A    &1.237&20.47&   21.35& 1&2.6&1.4$\pm$0.6&\ha &0.89& 91$\pm$49& 6$\pm$3\\
Q1623+268B    &1.237&19.96&   20.98& 2&4.5&2.5$\pm$0.6&\ha &0.89&115$\pm$37&10$\pm$3\\
Q1623+268C    &1.237&20.24&$>$22.18& 2&4.6&2.7$\pm$0.3&\ha &0.89& $>$248 & 8--14\\
Q1623+268D    &1.237&20.40&   21.29& 3&2.8&1.5$\pm$0.6&\ha &0.89& 92$\pm$47&6$\pm$3\\
Q1623+268E    &1.237&20.42&$>$22.18& 3&3.7&2.1$\pm$0.6&\ha &0.89& $>$180 &6--12\\
Q1623+268F    &1.237&20.50&   21.40& 3&2.6&1.4$\pm$0.6&\ha &0.89& 94$\pm$51&6$\pm$3\\
\hline
Q2038$-$012A  &2.248&18.09&   18.77& 2&3.5&1.7$\pm$0.4&\ha &2.42&37$\pm$10&68$\pm$16\\
Q2038$-$012B  &2.248&19.01&   20.73& 2&3.3&1.7$\pm$0.4&\ha &2.42&256$\pm$170&65$\pm$16\\
Q2038$-$012C  &2.248&18.44&   19.66& 1&4.5&2.3$\pm$0.4&\ha &2.42&  $115\pm$32&88$\pm$16\\
Q2038$-$012D  &2.248&18.90&$>$20.84& 1&3.9&2.0$\pm$0.4&\ha &2.42&$>$194 &60--94\\
\hline
Q2348$-$011A\tablenotemark{b}&2.248&18.74&   20.27& 2&3.0&2.0$\pm$0.6&\ha &2.43&  $191\pm$95&78$\pm$21\\
Q2348$-$011B  &2.248&18.81&$>$20.79& 3&3.2&2.2$\pm$0.6&\ha &2.43&$>$199&63--107\\
\tablenotetext{a}{R rank degree of significance: 1 is highest, 3 is lowest; 
  was estimated by inspection to take into account systematic uncertainties such
  as a very bright continuum, proximity to the edge of the field, and
  proximity of nearby objects making accurate magnitudes difficult to
  derive}
\tablenotetext{b}{Magnitudes may be affected by nearby objects}
\enddata
\end{deluxetable}

\end{document}